\def\version{}     



\font\bigbold=cmbx12
\font\ninerm=cmr9      \font\eightrm=cmr8    \font\sixrm=cmr6
\font\fiverm=cmr5
\font\ninebf=cmbx9     \font\eightbf=cmbx8   \font\sixbf=cmbx6
\font\fivebf=cmbx5
\font\ninei=cmmi9      \skewchar\ninei='177  \font\eighti=cmmi8
\skewchar\eighti='177  \font\sixi=cmmi6      \skewchar\sixi='177
\font\fivei=cmmi5
\font\ninesy=cmsy9     \skewchar\ninesy='60  \font\eightsy=cmsy8
\skewchar\eightsy='60  \font\sixsy=cmsy6     \skewchar\sixsy='60
\font\fivesy=cmsy5     \font\nineit=cmti9    \font\eightit=cmti8
\font\ninesl=cmsl9     \font\eightsl=cmsl8
\font\ninett=cmtt9     \font\eighttt=cmtt8
\font\tenfrak=eufm10   \font\ninefrak=eufm9  \font\eightfrak=eufm8
\font\sevenfrak=eufm7  \font\fivefrak=eufm5
\font\tenbb=msbm10     \font\ninebb=msbm9    \font\eightbb=msbm8
\font\sevenbb=msbm7    \font\fivebb=msbm5
\font\tenssf=cmss10    \font\ninessf=cmss9   \font\eightssf=cmss8
\font\tensmc=cmcsc10

\newfam\bbfam   \textfont\bbfam=\tenbb \scriptfont\bbfam=\sevenbb
\scriptscriptfont\bbfam=\fivebb  \def\Bbb{\fam\bbfam}
\newfam\frakfam  \textfont\frakfam=\tenfrak \scriptfont\frakfam=%
\sevenfrak \scriptscriptfont\frakfam=\fivefrak  \def\frak{\fam\frakfam}
\newfam\ssffam  \textfont\ssffam=\tenssf \scriptfont\ssffam=\ninessf
\scriptscriptfont\ssffam=\eightssf  
\def\smc{\tensmc}

\def\eightpoint{\textfont0=\eightrm \scriptfont0=\sixrm
\scriptscriptfont0=\fiverm  \def\rm{\fam0\eightrm}%
\textfont1=\eighti \scriptfont1=\sixi \scriptscriptfont1=\fivei
\def\oldstyle{\fam1\eighti}\textfont2=\eightsy
\scriptfont2=\sixsy \scriptscriptfont2=\fivesy
\textfont\itfam=\eightit         \def\it{\fam\itfam\eightit}%
\textfont\slfam=\eightsl         \def\sl{\fam\slfam\eightsl}%
\textfont\ttfam=\eighttt         \def\tt{\fam\ttfam\eighttt}%
\textfont\frakfam=\eightfrak     \def\frak{\fam\frakfam\eightfrak}%
\textfont\bbfam=\eightbb         \def\Bbb{\fam\bbfam\eightbb}%
\textfont\bffam=\eightbf         \scriptfont\bffam=\sixbf
\scriptscriptfont\bffam=\fivebf  \def\bf{\fam\bffam\eightbf}%
\abovedisplayskip=9pt plus 2pt minus 6pt   \belowdisplayskip=%
\abovedisplayskip  \abovedisplayshortskip=0pt plus 2pt
\belowdisplayshortskip=5pt plus2pt minus 3pt  \smallskipamount=%
2pt plus 1pt minus 1pt  \medskipamount=4pt plus 2pt minus 2pt
\bigskipamount=9pt plus4pt minus 4pt  \setbox\strutbox=%
\hbox{\vrule height 7pt depth 2pt width 0pt}%
\normalbaselineskip=9pt \normalbaselines \rm}

\def\ninepoint{\textfont0=\ninerm \scriptfont0=\sixrm
\scriptscriptfont0=\fiverm  \def\rm{\fam0\ninerm}\textfont1=\ninei
\scriptfont1=\sixi \scriptscriptfont1=\fivei \def\oldstyle%
{\fam1\ninei}\textfont2=\ninesy \scriptfont2=\sixsy
\scriptscriptfont2=\fivesy
\textfont\itfam=\nineit          \def\it{\fam\itfam\nineit}%
\textfont\slfam=\ninesl          \def\sl{\fam\slfam\ninesl}%
\textfont\ttfam=\ninett          \def\tt{\fam\ttfam\ninett}%
\textfont\frakfam=\ninefrak      \def\frak{\fam\frakfam\ninefrak}%
\textfont\bbfam=\ninebb          \def\Bbb{\fam\bbfam\ninebb}%
\textfont\bffam=\ninebf          \scriptfont\bffam=\sixbf
\scriptscriptfont\bffam=\fivebf  \def\bf{\fam\bffam\ninebf}%
\abovedisplayskip=10pt plus 2pt minus 6pt \belowdisplayskip=%
\abovedisplayskip  \abovedisplayshortskip=0pt plus 2pt
\belowdisplayshortskip=5pt plus2pt minus 3pt  \smallskipamount=%
2pt plus 1pt minus 1pt  \medskipamount=4pt plus 2pt minus 2pt
\bigskipamount=10pt plus4pt minus 4pt  \setbox\strutbox=%
\hbox{\vrule height 7pt depth 2pt width 0pt}%
\normalbaselineskip=10pt \normalbaselines \rm}

\global\newcount\secno \global\secno=0 \global\newcount\meqno
\global\meqno=1 \global\newcount\appno \global\appno=0
\newwrite\eqmac \def\romappno{\ifcase\appno\or A\or B\or C\or D\or
E\or F\or G\or H\or I\or J\or K\or L\or M\or N\or O\or P\or Q\or
R\or S\or T\or U\or V\or W\or X\or Y\or Z\fi}
\def\eqn#1{ \ifnum\secno>0 \eqno(\the\secno.\the\meqno)
\xdef#1{\the\secno.\the\meqno} \else\ifnum\appno>0
\eqno({\rm\romappno}.\the\meqno)\xdef#1{{\rm\romappno}.\the\meqno}
\else \eqno(\the\meqno)\xdef#1{\the\meqno} \fi \fi
\global\advance\meqno by1 }

\global\newcount\refno \global\refno=1 \newwrite\reffile
\newwrite\refmac \newlinechar=`\^^J \def\ref#1#2%
{\the\refno\nref#1{#2}} \def\nref#1#2{\xdef#1{\the\refno}
\ifnum\refno=1\immediate\openout\reffile=refs.tmp\fi
\immediate\write\reffile{\noexpand\item{[\noexpand#1]\ }#2\noexpand%
\nobreak.} \immediate\write\refmac{\def\noexpand#1{\the\refno}}
\global\advance\refno by1} \def\semi{;\hfil\noexpand\break ^^J}
\def\nl{\hfil\noexpand\break ^^J} \def\refn#1#2{\nref#1{#2}}

\newif\iftitlepage \titlepagetrue \newtoks\titlepagefoot
\titlepagefoot={\hfil} \newtoks\otherpagesfoot \otherpagesfoot=%
{\hfil\tenrm\folio\hfil} \footline={\iftitlepage\the\titlepagefoot%
\global\titlepagefalse \else\the\otherpagesfoot\fi}

\def\abstract#1{{\parindent=30pt\narrower\noindent\ninepoint\openup
2pt #1\par}}

\newcount\notenumber\notenumber=1 \def\note#1
{\unskip\footnote{$^{\the\notenumber}$} {\eightpoint\openup 1pt #1}
\global\advance\notenumber by 1}

\def\today{\ifcase\month\or January\or February\or March\or
April\or May\or June\or July\or August\or September\or October\or
November\or December\fi \space\number\day, \number\year}

\def\pagewidth#1{\hsize= #1}  \def\pageheight#1{\vsize= #1}
\def\hcorrection#1{\advance\hoffset by #1}
\def\vcorrection#1{\advance\voffset by #1}

\font\extra=cmss10 scaled \magstep0  \setbox1 = \hbox{{{\extra R}}}
\setbox2 = \hbox{{{\extra I}}}       \setbox3 = \hbox{{{\extra C}}}
\setbox4 = \hbox{{{\extra Z}}}       \setbox5 = \hbox{{{\extra N}}}





\def\frac#1#2{{#1\over#2}}

\def\pmb#1{\setbox0=\hbox{$#1$} \kern-.025em\copy0\kern-\wd0
    \kern.05em\copy0\kern-\wd0 \kern-.025em\raise.0433em\box0 }

\def\ve{\vfill\eject}

\def\Z{{\Zed}}
\def\R{{\Real}\!}

\def\({\left(}
\def\){\right)}
\def\[{\left[}
\def\]{\right]}


\def\R{{\rm I \hskip-.47ex R}}
\def\X{{\rm X \hskip-1.4ex X}}

\def\R{{\Bbb R}}  \def\X{{\Bbb X}}  \def\Z{{\Bbb Z}}  
\def\I{{\Bbb I}}

\def\Line{{\X}}

\def\PL{^{(+)}}     \def\PM{^{(\pm)}}     \def\ONE{^{(1)}}
\def\MI{^{(-)}}          \def\TWO{^{(2)}}

\def\text#1{{   \rm     #1    }}   
     
\def\texttt#1{{ \rm     #1 \  }}

\def\d{\text{d}}

\def\lhs{l.h.s.\ }
\def\rhs{r.h.s.\ }

\def\f#1#2{{#1\over#2}}
\def\ff#1#2{\raise.2pt\hbox{\ninepoint${\displaystyle\f{#1}{#2}}$}}
\def\fff#1#2{\raise.5pt\hbox{\eightpoint${\displaystyle\f{#1}{#2}}$}}

\def\biT{\hskip -0.15ex}

\def\bit{\hskip 0.15ex}
\def\bitt{\hskip 0.30ex}

\def\lc{limit-circle }    
\def\lp{limit-point }


{

\refn\Nature
{A. Fuhrer, S. L{\"u}sher, T. Ihn, T. Heinzel, K. Ensslin, W. Wegscheider
and M. Bichler,
{\it Nature} {\bf 413} (2001) 822}

\refn\CFT
{T. Cheon, T. F\"{u}l\"{o}p and I. Tsutsui,
{\it Ann.\ Phys.} {\bf 294} (2001) 1}

\refn\MT
{H. Miyazaki and I. Tsutsui,
{\it Ann. Phys.} {\bf 299} (2002) 78}

\refn\GC
{A.N. Gordeyev and S.C. Chhajlany, {\it J.\ Phys.\ A} {\bf 30}
(1997) 6893}

\refn\Lathouwers
{L. Lathouwers, {\it J.\ Math.\ Phys.} {\bf 16} (1975) 1393}

\refn\RS
{M. Reed and B. Simon,
\lq\lq Methods of Modern
Mathematical Physics II, Fourier analysis, self-adjointness\rq\rq,
Academic Press, New York, 1975}

\refn\AG
{N.I. Akhiezer and I.M. Glazman,
\lq\lq Theory of Linear Operators in Hilbert Space\rq\rq,
Vol.II,
Pitman Advanced Publishing Program, Boston, 1981}

\refn\BFV
{G. Bonneau, J. Faraut and G. Valent,
{\it Self-adjoint extensions of operators and the teaching of quantum
mechanics}, LPTHE preprint PAR/LPTHE/99-43, quant-ph/0103153}

\refn\FT
{T. F\"{u}l\"{o}p and I. Tsutsui,
{\it Phys.\ Lett.} {\bf A264} (2000) 366}

\refn\Krall
{A.M. Krall, {\it J.\ Differential Equations} {\bf 45} (1982) 128}

\refn\AGHH
{S. Albeverio, F. Gesztesy, R. H{\o}egh-Krohn and H. Holden,
\lq\lq Solvable Models in Quantum Mechanics\rq\rq,
Springer, New York, 1988}

\refn\Rellich
{F. Rellich, {\it Math.\ Ann.} {\bf 122} (1951) 343}

\refn\Fulton
{C.T. Fulton, {\it Trans.\ Amer.\ Math.\ Soc.} {\bf 229} (1977) 51;
{\it Proc.\ Roy.\ Soc.\ Eddingburgh Sect A} {\bf 87} (1980) 1}

\refn\Kochubei
{A.N. Kochubei, {\it Siberian Math.\ J.} {\bf 32}
(1991) 401}

\refn\Loudon
{R. Loudon, {\it Amer.\ J.\ Phys.} {\bf 27}
(1959) 649}

\refn\FLM
{W. Fisher, H. Leschke and P. M\"{u}ller,
{\it J.\ Math.\ Phys.} {\bf 36} (1995) 2313}

\refn\Geszt
{F. Gesztesy, {\it J.\ Phys.\ A} {\bf 13}
(1980) 867}

\refn\Moebius
{I. Tsutsui, T. F\"{u}l\"{o}p and T. Cheon,
{\it J.\ Math.\ Phys.} {\bf 42} (2001) 5687}


\refn\Calogero
{F. Calogero, {\it J.\ Math.\ Phys.} {\bf 10} (1969) 2191,
2197; {\bf 12} (1971) 419}

\refn\KW
{P. Kurasov and K. Watanabe,
{\it On Rank One $H_{-3}$-Perturbations of Positive
Self-adjoint Operators}, Stockholm Univ. Res. Rep. Math. 1999/10;
{\it On $H_{-4}$-perturbations of self-adjoint operators},
Stockholm Univ. Res. Rep. Math. 2000/16}

}


\pageheight{23cm}
\pagewidth{14.8cm}
\hcorrection{0mm}
\vcorrection{2mm}
\magnification= \magstep1
\def\bsk{%
\baselineskip= 16.8pt plus 1pt minus 1pt}
\parskip=5pt plus 1pt minus 1pt
\tolerance 6000




\version \hfill 
KEK Preprint 2002-95

\vskip 42pt

{\baselineskip=18pt

\centerline{\bigbold
Connection Conditions and the Spectral Family}
\centerline{\bigbold
under Singular Potentials}

\vskip 30pt

\centerline{\smc
Izumi Tsutsui\footnote{${}^*$}
{\eightpoint email:\quad izumi.tsutsui@kek.jp}
\quad
{\rm and}
\quad
Tam\'{a}s F\"{u}l\"{o}p\footnote{${}^\dagger$}
{\eightpoint email:\quad fulopt@poe.elte.hu}
}

\vskip 5pt

{
\baselineskip=13pt
\centerline{\it
Institute of Particle and Nuclear Studies}
\centerline{\it
High Energy Accelerator Research Organization (KEK)}
\centerline{\it
Tsukuba 305-0801, Japan}
}

\vskip 3pt
\centerline{\rm and}
\vskip 3pt

\centerline{\smc
Taksu Cheon\footnote{${}^\ddagger$}
{\eightpoint email:\quad cheon@mech.kochi-tech.ac.jp,
http://www.mech.kochi-tech.ac.jp/cheon/}
}

\vskip 3pt
{
\baselineskip=13pt
\centerline{\it Laboratory of Physics}
\centerline{\it Kochi University of Technology}
\centerline{\it Tosa Yamada, Kochi 782-8502, Japan}
}

\vskip 30pt

\abstract{%
{\bf Abstract.}\quad
To describe a quantum system whose potential is
divergent at one point, one must provide proper connection
conditions for the wave functions at
the singularity.
Generalizing the scheme used for
point interactions in one dimension, we present a set of connection
conditions which are well-defined even if the wave functions and/or their
derivatives are divergent at the singularity.  Our generalized scheme covers
the entire $U(2)$ family of quantizations (self-adjoint
Hamiltonians) admitted for the singular system.
We use this scheme
to examine the spectra of the Coulomb potential
$V(x) = - e^2/ \vert x\vert$
and the harmonic oscillator with square inverse potential
$V(x) = ({m\omega^2}/{2}) \,{x^2} + g/{x^2}$, and thereby provide a
general perspective for these models which have previously been treated with
restrictive connection conditions resulting in conflicting spectra.
We further show that, for any parity invariant singular potentials
$V(-x) = V(x)$, the spectrum is determined solely
by the eigenvalues of the characteristic matrix
$U \in U(2)$.
}

\vskip 10pt
%
%
%
%
}
\ve


\pageheight{23cm}
\pagewidth{15.7cm}
\hcorrection{-1mm}
\magnification= \magstep1
\def\bsk{%
\baselineskip= 14.8pt plus 1pt minus 1pt}
\parskip=5pt plus 1pt minus 1pt
\tolerance 8000
\bsk



\bigskip
\centerline
{\bf 1. Introduction}
\bigskip

Quantum singularity is a source of interesting physics and, at the same time,
confusion.  Even in its simplest form as a point interaction ---
now realized
approximately as quantum dots (see, {\it e.g.}, [\Nature]) ---
it provides unexpectedly rich quantum phenomena
such as duality and anholonomy [\CFT].
When it arises as a divergent point of an infinite potential
wall, it may admit quantum tunnelling through the infinite
wall and, in some cases, can lead to an exotic quantum
caustic [\MT].  However,
it also poses the problem in its own treatment
in quantum mechanics.   In fact, if we look back the history of the one
dimensional Coulomb potential,
$
V(x) = - e^2/ \vert x\vert
$, for instance, we find persistent
disagreement over the possible spectrum for
nearly a half century [\GC].   A similar
confusion can be found for a system with the square inverse potential,
$
V(x) = g/ x^2
$ [\Lathouwers].

These confusing circumstances arise due to the ambiguity in choosing
boundary (or connection) conditions at the singularity, which in mathematical
terms corresponds
to the choice of self-adjoint domains for the
Hamiltonian operator,
$$
H = - {{\hbar^2}\over{2m}}{{\d^2}\over{\d x^2}} + V(x).
\eqn\hamilt
$$
It has been known that a point interaction on a one dimensional line
admits a
$U(2)$ family of self-adjoint extensions for the Hamiltonian, and that these
are characterized by distinct connection conditions [\RS].  For a
point interaction occurring at $x = 0$, the connection conditions can be
given by
$$
(U-I)\Psi+i {L_0}(U+I)\Psi' =0,
\eqn\BoundaryCondition
$$
where $U \in U(2)$ is a matrix characterizing the self-adjoint extension,
$I$ is the identity matrix, and
$L_0 \ne 0$ is a constant with
dimension of length [\AG,\BFV,\FT].  $\Psi$ and $\Psi'$ are
boundary vectors
$$
\Psi =
  \left( {\matrix{{\psi (+0)}\cr
                  {\psi (-0)}\cr}
         }
  \right),
\qquad
\Psi' =
  \left( {\matrix{{ \psi' (+0)}\cr
                  {-\psi' (-0)}\cr}
         }
  \right),
\eqn\vectors
$$
defined from the boundary values $\psi (\pm 0) = \lim_{x \to \pm 0}
\psi(x)$ of the wave function $\psi$ and its derivative $\psi'\equiv
{{\d\psi}\over{\d x}}$.
The problem, however, is that this prescription of connection
conditions may not be directly applicable
to singular potentials $V(x)$, because then the
boundary values $\psi (\pm 0)$ and/or $\psi' (\pm 0)$ may diverge at the
singularity
and, accordingly, the vectors $\Psi$ and $\Psi'$ in (\vectors) become
ill-defined.

One of the purposes of the present paper is to
provide a scheme in which the
connection conditions (\BoundaryCondition) become well-defined
even for singular
potentials, and thereby furnish a general framework for studying
singular
systems on a line including those mentioned above.  Typically,
at the singularity
such systems
allow for two independent square integrable solutions
$x = 0$ for the eigenvalue equation $H \psi = E \psi$ for any $E$,
and (at most) only one square integrable solution at
$x \to \pm\infty$.  Systems of this type are said to be in the
{\it \lc case} at $x = 0$ and in the {\it \lp case} at infinity [\RS].
At a \lp singularity, no boundary condition is needed for ensuring the
self-adjointness of the Hamiltonian [\AG], while at a \lc singularity,
some boundary conditions are necessary in order to specify a self-adjoint
Hamiltonian from the family of possible self-adjoint domains. This family
is $U(2)$ for these systems, which follows, for example, from the fact
that each of the negative and the positive half lines has
one square integrable eigenmode for any nonreal
eigenvalue $E$ [\RS,\AG].
An essential step toward the generalization of the connection
conditions for such systems
consists of replacing the boundary values of the wave functions with
corresponding Wronskians.  This idea has been proposed [\Krall] for systems
on a
half line with a singular endpoint for which a $U(1)$ family
of boundary conditions is assigned. Here, we extend this to the full line,
where now the family is given by $U(2)$, in such a way that the
connection
conditions (\BoundaryCondition) remain to be valid with modified boundary
vectors.  (For a different scheme of providing the domains of possible
self-adjoint Hamiltonians, see [\AGHH].)

{}For illustration, we employ our scheme to analyze two models, one with the
Coulomb potential and the other with
the harmonic oscillator with square inverse potential
$V(x) = ({m\omega^2}/{2}) \,{x^2} + g/{x^2}$.
We shall see that the various different quantizations discussed
previously for those models arise at different choices of the matrix $U$,
and that the
spectra are dependent on the choice of $U$; in fact,
this dependence has caused the confusion
concerning the spectrum in the literature.
Interestingly, the spectrum depends on two parameters (the eigenvalues of
the matrix $U$), not all the four of
$U \in U(2)$, and this two-parameter dependence of the spectra is
shared by any parity invariant potentials
$V(-x) = V(x)$ with singularity of the kind just mentioned.  More precisely,
we find that for those systems the space of spectra is given by the
M{\"o}bius
strip
$U(1) \times U(1)/\Z_2$.

The plan of this paper follows the line of
arguments stated here,  that is, we give the generalized connection
conditions in
Sect.\bitt 2 and thereby analyze the two models in Sect.\bitt 3.
The two-parameter dependence of the
spectra is then established in Sect.\bitt 4, and finally Sect.\bitt 5 is
devoted to summary and discussions.

\vfill\eject
\centerline
{\bf 2. Connection conditions}
\bigskip

To begin with, we provide the general connection conditions by extending the
construction proposed in Ref.~[\Krall] (see also
[\Rellich,\Fulton,\Kochubei]) from the half line to the full line.
Let the potential $V$ possess a singularity at $x = 0$ on the one dimensional
line $\Line$.\note{%
We use the symbol $\Line$ to stress that it is dimensionful in contrast to
the  dimensionless real line
$\R$.}
The potential is assumed to be in the \lc case at $x = 0$ from both sides and
in the \lp case at $x \to
\pm
\infty$, and regular otherwise.
We first consider the maximum domain ${\cal F} \subset L^2(\Line)$ on which the
Hamiltonian $H$ can be defined as a differential operator,
$$
\eqalign{
{\cal F} = \big\{ \, \psi \in L^2 (\Line & ) \ \big| \ \hbox{$\psi$
and $\psi'$ are absolutely continuous on} \cr & \hbox{every finite
subinterval of $\, \Line \! \setminus \! \{ 0\} $} \, ,
\ \ H \psi \in L^2(\Line) \, \big\} \, .
}
\eqn\fdefnit
$$
The Hamiltonian is not symmetric on ${\cal F}$, since, for $\phi$,
$\psi \in {\cal F}$, we have
$$
\int_\Line \d x \[ \phi^* H \psi -  (H \phi)^* \psi \] =
\ff{\hbar^2}{2m} \( W[\phi^*, \psi]_{+0} - W[\phi^*, \psi]_{-0} \),
\eqn\fntval
$$
where $W[\phi^*,\psi]_{\pm 0}$ are the limiting values for $x \to \pm 0$ of
the Wronskian
$$
W[\phi^*,\psi](x) = \phi^*(x) \bit \psi'(x)-{\phi^*}'(x) \bit \psi(x).
\eqn\wronsk
$$
Here we have utilized the facts that $W[\phi^*, \psi]$ vanishes for $x \to
\pm \infty$, since the infinites are \lp [\AG], and that it is finite
in the limits $x \to \pm 0$ even if the two
functions $\phi(x)$, $\psi(x)$ are
divergent. This latter can be shown as follows. For $\epsilon > 0$, we
introduce the space of functions
$$
\eqalign{
{\cal F}_{\epsilon} = \big\{ \, \psi \in L^2 (0 & , \epsilon ) \ \big| \
\hbox{$\psi$ and $\psi'$ are absolutely} \cr & \hbox{continuous on }
(0, \epsilon)\bitt , \ \ H \psi \in L^2 (0, \epsilon) \, \big\} \, .
}
\eqn\fedefnit
$$
Note that ${\cal F}_{\epsilon}$ contains ${\cal F}$ as well as a wide
range of other interesting functions as well, including {\it all} the
eigenfunctions of the differential operator $H$, which are square
integrable in any finite neighbourhood of the limit-circle singularity $x
= 0$ but not necessarily on the whole line $\Line$. Now, for $\phi, \psi
\in {\cal F}_{\epsilon}$ and $0 < \delta < \epsilon$,
$$
\int_\delta^\epsilon \d x \[ \phi^* H \psi -  (H \phi)^* \psi \] =
\ff{\hbar^2}{2m} \( W[\phi^*, \psi]_{\delta} -
W[\phi^*, \psi]_{\epsilon} \).
\eqn\ffntval
$$
Both terms of the \rhs are finite. The \lhs is also finite, even if we
let $\delta \to 0$. Consequently, $ \lim_{\delta \to 0} W[\phi^*,
\psi]_{\delta} = W[\phi^*, \psi]_{+0}$ is finite. As one can see, this
property holds actually not only in ${\cal F}$ but even in ${\cal
F}_{\epsilon}$. The finiteness of $W[\phi^*, \psi]_{-0}$ is proved
similarly, with the aid of the analogously introduced ${\cal
F}_{-\epsilon}$.

Since the \rhs of (\fntval) is generally nonvanishing, any self-adjoint
domain ${\cal D}$ for $H$ must be such a subset of ${\cal F}$ that the
\rhs of (\fntval) is zero for all functions within ${\cal D}$. Now we show
how to characterize the possible self-adjoint domains via a connection
condition at $x = 0$ of the form (\BoundaryCondition), where the boundary
vectors $\Psi$, $\Psi'$ are appropriately generalized
with the help of a basic set of energy eigenmodes.
Let $\varphi^{(i)}$, for $ i = 1$, 2, be two independent, {real}
eigenmodes with eigenvalue $E$,
$$
H\varphi^{(i)}(x) = E\, \varphi^{(i)}(x),
\qquad
W[\varphi^{(1)},\varphi^{(2)}](x) = 1,
\eqn\zeromode
$$
for $x \ne 0$. The actual value of $E$ is unimportant for our purposes.
We note that these eigenmodes may not be square integrable on the whole
line and hence may not belong to ${\cal F}$, but they necessarily belong
to ${\cal F}_{\epsilon}$ and ${\cal F}_{-\epsilon}$.
Consequently, the complex column vectors
$$
\Psi=\pmatrix{
W[\psi,\varphi^{(1)}]_{+0}\cr
W[\psi,\varphi^{(1)}]_{-0}
},
\qquad
\Psi'=\pmatrix{
W[\psi,\varphi^{(2)}]_{+0}\cr
- W[\psi,\varphi^{(2)}]_{-0}
}
\eqn\boundvect
$$
are well-defined for $\psi \in {\cal F}$
since they are constructed from finite quantities.
Further, observing that we can rewrite the Wronskian (\wronsk) as
$$
\eqalign{
W[\phi^*,\psi]
&=
\left\vert
\matrix{
\phi^* & {\phi^*}'\cr
\psi   & \psi' \cr
}
\right\vert_{{}_{}}
=
\left\vert
\matrix{
\phi^* & {\phi^*}'\cr
\psi   & \psi' \cr
}
\right\vert
\left\vert
\matrix{
{\varphi\ONE}' & {\varphi\TWO}'\cr
-{\varphi\ONE} & -{\varphi\TWO} \cr
}
\right\vert_{{}_{}} \cr
&=
\left\vert
\matrix{
\phi^*{\varphi\ONE}' - {\phi^*}'\varphi\ONE & \phi^*{\varphi\TWO}' -
{\phi^*}'\varphi\TWO\cr
\psi{\varphi\ONE}' - \psi'\varphi\ONE & \psi{\varphi\TWO}' -
\psi'\varphi\TWO \cr } \right\vert \cr
&=
W[\phi^*, \varphi\ONE] \bitt W[\psi, \varphi\TWO] - W[\phi^*, \varphi\TWO]
\bitt W[\psi, \varphi\ONE] \, ,
}
\eqn\newwronsk
$$
we can express the \rhs of (\fntval) in terms of the boundary vectors
for
$\phi$ and $\psi$ simply as
$$
\ff{\hbar^2}{2m} \[ \Phi^\dagger \Psi' - \Phi'^\dagger \Psi \] ,
\eqn\aaaa
$$
where $\Phi$ and $\Phi'$ are introduced for $\phi$ analogously to (\boundvect) for
$\psi$.  If $\phi$, $\psi$ are in a self-adjoint domain ${\cal D}$ then (\aaaa)
must vanish. In particular, for $\phi = \psi$, this condition reads
$\Psi^\dagger \Psi' = \Psi'^\dagger \Psi$, which, under the notations
$$
\Psi\PM = \Psi \pm i L_0 \Psi'
\eqn\aaab
$$
with an arbitrary nonzero constant $L_0$, is
equivalent to the equality of the norms
$ \| \Psi\PL \| = \| \Psi\MI \| $.  This shows
that
$\Psi\PL$ and $\Psi\MI$ are in a relationship
$$
U \Psi\PL = \Psi\MI \, , \hskip 10ex U \in U(2) \, ,
\eqn\aaad
$$
which is nothing but (\BoundaryCondition). Different states $\phi$, $\psi
\in {\cal D}$ have to share the same $U$ so as to make (\aaaa) identically
vanish:
$$
\eqalign{
\ff{\hbar^2}{2m} \[ \Phi^\dagger \Psi' - \Phi'^\dagger \Psi \] & =
\ff{\hbar^2}{4imL_0} \bit \big[ \bit {\Phi\PL}^\dagger \Psi\PL -
{\Phi\MI}^\dagger \Psi\MI \bit \big] \cr & =
\ff{\hbar^2}{4imL_0} \bit \big[ \bit {\Phi\PL}^\dagger \Psi\PL -
\big( U \Phi\PL \big)^\dagger \big( U \Psi\PL \big) \big] = 0 \, .  }
\eqn\aaac
$$
By an argument analogous to the case of the half line [\Krall], it is not
hard to show that the connection condition (\BoundaryCondition) restricts
the space ${\cal F}$ to a domain ${\cal D} \equiv {\cal D}_U$ on which the
Hamiltonian is not only symmetric but indeed self-adjoint. Since all
different $U$s specify different self-adjoint domains ${\cal D}_U$, here
we can see again that, as in the case of point interactions, the family of
self-adjoint Hamiltonians $H \equiv H_U$ allowed on the line with
potential $V(x)$ possessing a \lc singularity and \lp behavior for $x \to
\pm \infty$ is given by $U(2)$. $U$ will be called the {\it
characteristic matrix} of the self-adjoint Hamiltonian $H_U$.

Various subfamilies of $U(2)$ can be defined analogously to the case of point
interactions.  For instance, the \lq separated subfamily\rq{} $\Omega_3$
where no probability flow through $x = 0$ is allowed is characterized by
those $U$ which are diagonal.  Indeed, for diagonal $U$ the probability
current
$$
j(x) =
\ff{\hbar}{2im} ( \psi^* \psi' -  \psi'^* \psi ) =
\ff{\hbar}{2im} \left(
W[\psi^*, \varphi\ONE] W[\psi, \varphi\TWO] - W[\psi^*, \varphi\TWO]
W[\psi, \varphi\ONE] \right)
\eqn\no
$$
is seen to vanish at $x = 0$, and diagonal $U$s provide the cases when
the boundary condition (\aaad) does not
mix the $+0$ boundary values with the $-0$ ones.
Hence we have $\Omega_3 \simeq U(1) \times U(1) \subset U(2)$, which are,
in other words, the cases where the system consists of two
independent half line systems.

If the domain of a self-adjoint Hamiltonian contains only functions that
are regular at the singularity, then one may
choose as reference modes any basis of independent solutions satisfying
$$
\varphi^{(1)}(\pm 0) = 0, \qquad
{\varphi^{(1)}}'(\pm 0) = 1, \qquad
\varphi^{(2)}(\pm 0) = -1, \qquad
{\varphi^{(2)}}'(\pm 0) = 0.
\eqn\normalization
$$
Under this choice, we find that the boundary
vectors (\boundvect) reduce to the conventional
form (\vectors), which shows that
our connection conditions are a natural generalization of the
conventional conditions.  Once generalized, however, we recognize that the normalizations
(\normalization) are not at all essential in presenting the connection conditions
(\BoundaryCondition) at the singularity to ensure the self-adjointness of the
Hamiltonian.  This in turn suggests that the characteristic matrix $U$ characterizes the
singularity only with respect to the reference modes chosen, and this fact has been
implicit in the previous treatment for non-singular cases based on the normalizations
(\normalization) .

\vfill\eject
\centerline
{\bf 3. Two models with singular potential}
\bigskip

We now employ the scheme just presented to analyze the two models
mentioned in the Introduction.

\medskip
\noindent
{\bf (i) One dimensional hydrogen atom}

\noindent
The first model is the one dimensional hydrogen atom, which is governed by
 the Coulomb potential,
$$
V(x) = - \frac{e^2}{\vert x\vert}.
\eqn\coulomb
$$
This system has a long history of research, dating back to Loudon [\Loudon]
who
first gave a set of bound state solutions ($E_n < 0$) for
the Schr{\"o}dinger equation,
$$
H {\psi_n}(x) = {E_n}{\psi_n}(x),
\eqn\egneq
$$
in terms of the Whittaker functions.  The spectrum obtained in [\Loudon] is
$$
E_n = - \frac{me^4}{2\hbar^2 n^2},
\qquad n = 1, 2, \ldots,
\eqn\onedspec
$$
where each level is doubly degenerate.
The system has later been examined by a number of other groups to obtain
different
spectra due to different choices of the connection condition at the
singularity $x
= 0$ (see, {\it i.e.},  [\FLM, \GC] and references therein). The connection
condition adopted originally in [\Loudon] is the Dirichlet condition
$\psi(\pm 0) = 0$ which corresponds to the Friedrichs extension of the
Hamiltonian [\Geszt], but other extensions are equally possible as we
shall now describe.

To apply our scheme of connection conditions, we first recall that in
terms of the variables
$$
z = 2\eta x, \qquad \eta = {{\sqrt{-2m E_n}}\over\hbar},
\qquad \alpha = {{e^2}\over\hbar}\sqrt{{-m}\over{2E_n}}.
\eqn\vrbls
$$
the Schr{\"o}dinger equation (\egneq) becomes
$$
{{\d^2\psi_n}\over{\d z^2}}  + \left({\alpha\over{\vert z\vert}} -
\frac{1}{4} \right) \psi_n = 0.
\eqn\wtk
$$
This is a special case of Whittaker's differential equation, whose
two independent solutions are the regular Whittaker function,
$$
M_{\alpha, \frac{1}{2}}(z) = z\, e^{-\frac{z}{2}} \, F(1-\alpha, 2;  z),
\eqn\emdf
$$
where $F(\alpha,\gamma;z)$ is the confluent
hypergeometric function,
and the irregular one,
$$
\eqalign{
W_{\alpha, \frac{1}{2}}(z)
&= {{e^{-\frac{z}{2}}}\over{\Gamma(-\alpha)}}
\biggl\{
z F(1-\alpha, 2;  z)
\left[
{\rm ln}\,z + \psi(1-\alpha) - \psi(1) - \psi(2)
\right] \cr
&\qquad \qquad \quad - \frac{1}{\alpha} + \sum_{r = 1}^\infty
{{(1-\alpha)_r}\over{r! (r+1)!}}
A_r z^{r+1}
\biggr\} .
}
\eqn\no
$$
Here $\Gamma(x)$ is the Gamma function, $\psi(x) = {d\over{dx}}{\rm ln}\,
\Gamma(x)$ is the di-Gamma function, and
$$
A_r = \sum_{n = 0}^{r-1} \left[{1\over{n+1-\alpha}} - {1\over{n+1}} -
{1\over{n+2}}\right],
\qquad (c)_r = {{\Gamma(c + r)}\over{\Gamma(c)}}.
\eqn\no
$$
{}From the asymptotic behavior of the two solutions, one finds that
$W_{\alpha, \frac{1}{2}}(z) $ is square integrable whereas
$M_{\alpha, \frac{1}{2}}(z) $ is not.  Thus the bound state
must be of the form,
$$
\psi_n(x) = W_{\alpha, \frac{1}{2}}(\vert z\vert)
\left\{ N_R\, \Theta(x) + N_L\, \Theta(-x) \right\},
\eqn\bstate
$$
where $\Theta(x)$ is the Heaviside step function, and $N_R$ and $N_L$ are
constants to be determined by the connection condition at $x = 0$.  Note
that, since asymptotically
$$
W_{\alpha, \frac{1}{2}}(z) =
{1\over{\Gamma(-\alpha)}}\left\{ - {1\over\alpha} + z\left[{\rm ln}\,z +
\psi(1-\alpha) - \psi(1) - \psi(2)
\right] \right\} + {\cal O}(z^2 {\rm ln}\,z),
\eqn\asymp
$$
as $z \to 0$, the bound state $\psi_n(x)$ has finite limits at
$x \to \pm 0$ whereas the derivative $\psi_n'(x)$ diverges there.

To see which bound states are actually allowed by the connection condition
(\BoundaryCondition), let us first fix the reference modes $\varphi\ONE$,
$\varphi\TWO$, in conformity with (\zeromode).  We choose them as
$$
\eqalign{
\varphi^{(1)}(x)
&= {1\over{2\kappa}} M_{\beta, \frac{1}{2}}(2\kappa\vert
x\vert)\left[\Theta(x) -
\Theta(-x)\right],\cr
\varphi^{(2)}(x)
&= -
\Gamma(1-\beta)\, W_{\beta, \frac{1}{2}}(2\kappa\vert x\vert),
}
\eqn\noref
$$
with
$$
\kappa = {{\sqrt{-2m  E}}\over\hbar},
\qquad \beta = {{e^2}\over\hbar}\sqrt{{-m}\over{2 E}},
\eqn\vrblstwo
$$
which are analogs of (\vrbls) with $E_n$ replaced by some arbitrary
$E < 0$.  With these, the boundary vectors (\boundvect) become finite as
they ought to be, and they are proportional to each other,
$$
\Psi= \sigma
\pmatrix{
N_R\cr
N_L
},
\qquad
\Psi'= \xi
\Psi,
\eqn\boundve
$$
where
$$
\sigma = {{1}\over{\Gamma(1-\alpha)}}, \qquad
\xi = {{2me^2}\over{\hbar^2}}\left[
\ln{\alpha\over\beta} - \psi(1-\alpha) + \psi(1-\beta)
\right].
\eqn\coefs
$$

With these, the connection conditions (\BoundaryCondition) read
$$
\left[(U-I)+i{L_0}\xi(U+I)\right]\Psi=0.
\eqn\boundarycon
$$
For $\psi_n$ to be a nontrivial bound state, we need
$$
\det
\left[
U-I+i{L_0}\xi(U+I)
\right]
=\det\left[
D-I+i{L_0}\xi(D+I)
\right]
=0,
\eqn\BoundaryConditionTwo
$$
where $D$ is a diagonal
matrix appearing in the standard decomposition,
$$
U = V^{-1}DV, \qquad  V \in SU(2).
\eqn\sdecp
$$
In terms of the parameterization,
$$
D=
\pmatrix{
e^{i{\theta_{+}}} & 0\cr
0 & e^{i{\theta_{-}}}
}, \qquad \theta_\pm \in
[0, 2\pi),
\eqn\dipara
$$
and
$$
{L_\pm}={L_0}\cot
\left(
\frac{{\theta_\pm}}{2}
\right).
\eqn\sclp
$$
we find that (\BoundaryConditionTwo) is satisfied if
$$
\xi =-\frac{1}{L_+} \quad\hbox{or}\quad -\frac{1}{L_-}.
\eqn\spesol
$$
Thus, given the singularity specified by $U$, we can determine
the spectrum of the bound states as solutions of (\spesol).  We observe
that the spectrum depends only on the two angles $(\theta_+, \theta_-)$ in
the diagonal part of $U$, that is, the two eigenvalues of $U$.  Later we
show that this is in fact the case for all parity invariant potentials
$V(-x) = V(x)$ sharing the same singular property considered here.

A particularly simple spectrum is obtained at the angles
$(\theta_{+},\theta_{-})=(\pi,\pi)$, {\it i.e.}, at $U = -I$.  The
connection condition (\boundarycon) then implies
$\Psi = 0$, and therefore we need $\sigma = 0$. {}From (\coefs) we learn
that this is fulfilled for $\alpha = 1$, $2, \ldots$, reproducing the
spectrum (\onedspec) with double degeneracy (since $N_R$ and $N_L$ are
chosen freely modulo the normalization).  Note that, since
$\varphi\ONE(\pm 0) = 0$ and
${\varphi\ONE}'(\pm 0) = 1$, the condition $\Psi = 0$ is
equivalent to demanding $\psi(\pm 0) = 0$, which is obtained by the so
called Friedrichs extension discussed in [\Geszt]. We, however, emphasize
that the Friedrichs extension is a special self-adjoint extension
belonging to the separated subfamily $\Omega_3$ where the two half lines are
physically decoupled and none of the scale parameters $L_\pm$ in (\sclp)
appears in the spectrum, but any
other extensions whose spectrum varies with the
parameters
$(\theta_+, \theta_-)$ in $U$ through $L_\pm$ are
equally possible.\note{%
The four-parameter family of extensions for the one dimensional Coulomb
system has been argued in [\FLM] in a slightly different scheme.}

We next turn to the scattering phenomena of the Coulomb potential under
our general connection conditions.
(These considerations will be valid for the repulsive Coulomb force as
well, with setting $e^2 < 0$.) {}For this, in place of (\vrblstwo)
we consider a positive scattering energy $E_k > 0$ and use
$$
k={\sqrt{2mE_k} \over \hbar}\ ,
\qquad
\gamma =-{{e^2} \over \hbar} \sqrt{{m} \over {2E_k}}\ .
\eqn\scaa
$$
Then the two independent solutions are still given by the
Whittaker functions with the new $\alpha = i\gamma$ and $z = 2ik\vert
x\vert$. One may choose the following real combinations for a set of two
independent solutions:
$$
\eqalign{
\phi_{1}(x) &=
 {1\over{2ik}}
 M_{i\gamma,\frac{1}{2}}(2ik\vert x\vert)
\left[\Theta(x) - \Theta(-x)\right],
\cr
\phi_{2}(x) &=
-{1\over 2}\left\{
\Gamma(1-i\gamma)\,
W_{i\gamma,\frac{1}{2}}(2ik\vert x\vert)
+
\Gamma(1+i\gamma)\,
W_{-i\gamma,\frac{1}{2}}(-2ik\vert x\vert)
\right\}.
}
\eqn\scab
$$
In passing, we note that the phase part of $\Gamma(1 + i\gamma)$, {\it i.e.},
$\eta_0 =\arg\Gamma(1+i\gamma)$ is called
\lq s-wave Coulomb phase shift\rq{} while its
modulus can be evaluated as
$$
\vert \Gamma(1\pm i\gamma) \vert
= \sqrt{ \Gamma(1+ i\gamma) \Gamma(1- i\gamma)}
= \sqrt{ {2\pi\gamma} \over {e^{\pi\gamma}-e^{-\pi\gamma}} }.
\eqn\no
$$
The set (\scab) is chosen so that the solutions satisfy
$
W[\phi\ONE(x),\phi\TWO(x)]=1
$
and normalized as
$$
\phi_{1}(\pm 0) = 0, \qquad
\phi_{1}'(\pm 0) = 1, \qquad
\phi_{2}(\pm 0) = -1,
\eqn\no
$$
whereas $\phi_{2}'(\pm 0)$ are divergent and cannot be normalized.
{}From (\emdf) and (\asymp) the asymptotic forms of the solutions for $\vert x
\vert
\to \infty$ are found to be
$$
\eqalign{
\phi_{1}(x) &\sim
{{e^{{\pi\over 2}\gamma}}\over{\vert \Gamma(1+ i\gamma)\vert}}{1\over{k}}
       \sin{ (k\vert x\vert-\gamma \ln 2k\vert x\vert + \eta_0) }
       \left[\Theta(x) -\Theta(-x)\right],\cr
\phi_{2}(x) &\sim
-\vert \Gamma(1+ i\gamma)\vert
e^{-{{\pi}\over 2}\gamma}
      \cos{ (k\vert x\vert-\gamma \ln 2k\vert x\vert+\eta_0 ) }.
}
\eqn\scac
$$

The general solution for scattering states is a linear combination of the two
solutions,
$$
\psi(x) =
\left\{
N_R^{(1)} \phi_{1}(x) + N_R^{(2)} \phi_{2}(x)
\right\} \Theta(x)
+
\left\{
N_L^{(1)} \phi_{1}(x) + N_L^{(2)} \phi_{2}(x)
\right\} \Theta(-x).
\eqn\scad
$$
Now, from the asymptotic behaviors (\scac) one deduces that the incoming
wave from the left is the one that behaves for $x \pm \infty$ as
$$
\eqalign{
\psi(x)
&\sim  T\, e^{ i(k x-\gamma \log 2k x) } \hskip 123.5pt
(x\to + \infty) \cr
&\sim    e^{ i(k x+\gamma \ln (-2kx)) }
              + R\, e^{-i(k x+\gamma \ln (-2kx) ) }
    \ \ \ \ \ \ \ \ \
(x\to-\infty).
}
\eqn\scaf
$$
This corresponds to the choice
$$
N_R^{(1)}
= ik\, {e^{-{\pi\over 2}\gamma -i\eta_0} \vert \Gamma(1+ i\gamma)\vert}
\, T,
\qquad
N_R^{(2)}
= - {{e^{{\pi\over 2}\gamma -i\eta_0}}\over{\vert \Gamma(1+ i\gamma)\vert}}\,  T,
\eqn\no
$$
and
$$
N_L^{(1)}
= -ik\, {e^{-{\pi\over 2}\gamma -i\eta_0} \vert \Gamma(1+ i\gamma)\vert}
\, (R - e^{2i\eta_0}),
\qquad
N_L^{(2)}
= - {{e^{{\pi\over 2}\gamma -i\eta_0}}\over{\vert \Gamma(1+ i\gamma)\vert}}\,
(R + e^{2i\eta_0}).
\eqn\no
$$

To implement the connection condition at $x = 0$, we need to find a set of
reference modes, $\varphi\ONE(x)$ and $\varphi\TWO(x)$ satisfying
(\zeromode).  For this we shall use the same set (\scac) of the solutions,
$\varphi^{(i)}(x) = \phi_i(x)$ for $i = 1$, 2, with reference
energy $E$ and, accordingly, with 
the corresponding parameter $\gamma(E)$ obtained from (\scaa).  The
boundary vectors (\boundvect)  can then be evaluated as
$$
\Psi=
- \pmatrix{
N_R^{(2)}\cr
N_L^{(2)}
},
\qquad
\Psi'=
\pmatrix{
N_R^{(1)}\cr
- N_L^{(1)}
}
+ \rho
\pmatrix{
N_R^{(2)}\cr
N_L^{(2)}
},
\eqn\scae
$$
with
$$
\rho = {{me^2}\over{\hbar^2}}\left\{f(E) - f(E_k)\right\},
\qquad f(s) =
\left\{
2\ln{\gamma(s)}
- \psi(1-i\gamma(s)) - \psi(1 + i\gamma(s))\right\}.
\eqn\refdep
$$
Plugging the vectors (\scae) into the connection condition
(\BoundaryCondition),
and solving for the scattering matrix, one obtains
$$
\pmatrix{
T\cr
R}
=
{
  -e^{2i\eta_0}
\over
  { (U-I) - \omega L_0(U+I)
  }
}
\left[ (U-I) + \omega^* L_0(U+I) \right]
\pmatrix{
0\cr
1},
\eqn\scah
$$
where we have used
$$
\omega = k\, e^{-{\pi \over 2} \gamma} \vert \Gamma(1+
i\gamma)\vert^2 + i\rho\, e^{{\pi\over 2}\gamma}.
\eqn\cusfact
$$
Note that the Friedrichs extension $U = -I$ allows no transmission $T=0$.
In fact, as seen easily in (\scah), this is the case for any
diagonal
$U$, which is expected from the fact that those $U$ belong to the separated
subfamily
$\Omega_3$.

To render the scattering data (\scah) more explicit, we use
the decomposition (\sdecp) with $D$ given in (\dipara)
and $V$ parametrized as
$$
V = e^{i{\mu\over 2}\sigma_2} e^{i{\nu\over 2}\sigma_3} \ ,
\qquad \mu \in [0, \pi], \quad \nu \in
[0, 2\pi) \, .
\eqn\contv
$$
Then, in terms of the scale parameters (\sclp) and
$$
\chi_\pm = \arg(1 +i\omega L_\pm),
\eqn\no
$$
the scattering formula reads
$$
\pmatrix{
T\cr
R}
= -e^{i(2\eta_0 + \chi_+ + \chi_-)}
\pmatrix{
i\sin(\chi_+ - \chi_-)\sin\mu\, e^{-i\nu}
\cr
\cos(\chi_+ - \chi_-) - i\sin(\chi_+ - \chi_-)\cos\mu }.
\eqn\no
$$

We note that this outcome depends on the choice of the reference
modes, not only on the matrix $U$.  This is a consequence of the fact that
the combination of $U$ and the reference modes, not each, is essential for the
determination of the singularity, as mentioned earlier.  The ambiguity in the
choice of the reference modes is described by the group $SL(2, \R)$ on account of
the reality condition of the modes and the normalization in the Wronskian (\zeromode).
Precisely which combinations of the parameters are physically important is an
interesting question and will be discussed elsewhere.

\bigskip
\noindent
{\bf (ii) Harmonic oscillator with  inverse square potential}

Our second model is the harmonic oscillator with inverse square potential,
$$
V(x) = \frac{m {\omega^2}}{2} {x^2} +
g\frac{1}{x^2}.
\eqn\howispot
$$
In contrast to the previous example, we consider the repulsive case $g > 0$
to examine the positively divergent potential, and add the quadratic term
to render the entire spectrum discrete.  To comply with the condition that
the
singularity be of the \lc case at $x=0$, we confine ourselves to
$$
0 < g < \frac{3\hbar^2}{8m},
\eqn\gregion
$$
for which the Hamiltonian admits a $U(2)$ family of extensions.
To solve the Schr{\"o}dinger equation (\egneq), let us set
$$
\psi_n  (x)={y^{a + 1/2}} e^{-{y^2}/2}\,
{f_n}({y^2}), \qquad
y =\sqrt{\frac{m\omega}\hbar} \, x,
\eqn\Ansatz
$$
for $x > 0$, and use
$$
a=
{1\over 2}\sqrt{1+\frac{8mg}{\hbar^2}}, \qquad z = y^2,
\eqn\svalue
$$
so that (\egneq) becomes
$$
z \frac{\d^2\biT f_n}{\d z^2}(z)
+\left(a + 1 -z\right)\frac{\d \biT f_n}{\d z}(z)
-\frac{1}{2}\left(a + 1 -\lambda_n\right)
f_n(z)=0, \qquad
\lambda_n = \frac{E_n}{\hbar\omega}.
\eqn\chgeqation
$$
This is just the confluent
hypergeometric differential equation, and hence the two independent
solutions
for (\egneq) are
$$
\eqalign{
\phi^{(1)}_n(x) &:={y^{c_1 - 1/2}}e^{-{y^2}/{2}}
F\left(\frac{{c_1}-\lambda_n}2,{c_1};{y^2}\right), \qquad c_1 = 1 + a,\cr
{\phi^{(2)}_n}(x) &:={y^{c_2 - 1/2}}e^{-{y^2}/{2}}
F\left(\frac{{c_2}-\lambda_n}2,{c_2};{y^2}\right), \qquad c_2 = 1 - a.
}
\eqn\Solution
$$
Since the solution for $x < 0$ can be found by setting $x \to -x$
in (\Solution),
the general solution for the bound state is given by
$$
\psi_n (x)
=[{N_{\rm R}^{(1)}}\phi^{(1)}_n(\vert x\vert)+{N_{\rm
R}^{(2)}}\phi^{(2)}_n(\vert x\vert)]\Theta(x) +[{N_{\rm
L}^{(1)}}\phi^{(1)}_n(\vert x\vert)+{N_{\rm
L}^{(2)}}\phi^{(2)}_n(\vert x\vert)]\Theta(-x),
\eqn\solutions
$$
where the constants $N_{\rm R}^{(s)}$ and $N_{\rm L}^{(s)}$ will be
restricted by the connection condition.
Note that (\gregion) guarantees that both of the two solutions are square
integrable
near the singularity.  The entire square integrability is then ensured if
the
solutions vanish sufficiently fast at the infinity $x \to \pm \infty$.
{}From the asymptotic behavior of the confluent hypergeometric function,
$$
F(\alpha,\gamma;z)\approx\frac{\Gamma(\gamma)}
{\Gamma({\alpha})}{e^z}{z^{\alpha-\gamma}},\qquad
\hbox{as}\quad  |z| \rightarrow \infty,
\eqn\no
$$
the square integrability of the solutions (\Solution) implies
$$
\frac{N_{\rm R}^{(1)}}{N_{\rm R}^{(2)}}=\frac{N_{\rm L}^{(1)}}{N_{\rm
L}^{(2)}}=
-\frac{\Gamma\left(({c_1}-\lambda_n)/2\right)}
{\Gamma\left(({c_2}-\lambda_n)/2\right)}
\frac{\Gamma(c_2 )}{\Gamma(c_1)}.
\eqn\Ratio
$$

Now for the reference modes (\zeromode), we choose two eigenmodes
belonging to an arbitrarily fixed energy $E$, given in terms of the
solutions (\Solution), as
$$
\eqalign{
\varphi^{(1)}(x) &= \sqrt{\frac{\hbar}{m\omega}}\,
\phi^{(1)}(\vert x\vert)\left[\Theta(x) -
\Theta(-x)\right],\cr
\varphi^{(2)}(x) &=
\frac{1}{c_2  -  c_1}\,
\phi^{(2)}(\vert x\vert).
}
\eqn\no
$$
{}From
$F(\alpha,\gamma;z)= 1 + {\cal O}(z)$ as $z \to 0$,
the boundary vectors (\boundvect) are found to be
$$
\Psi= ({c_1}-{c_2})
\pmatrix{
N_{\rm R}^{(2)}\cr
N_{\rm L}^{(2)}
},
\qquad
\Psi'=\sqrt{\frac{m\omega}\hbar} \,
\pmatrix{
N_{\rm R}^{(1)}\cr
N_{\rm L}^{(1)}
},
\eqn\Vectors
$$
which are finite despite that the solution (\solutions) is divergent
at the singularity.  Moreover, we observe from
the relations
(\Ratio) and (\Vectors) that the two vectors
$\Psi'$ and $\Psi$ are again proportional to each other, and hence
if we write
$\Psi'=\xi \Psi$ the boundary condition (\BoundaryCondition)
becomes (\boundarycon) as before.  Combining (\Ratio) and (\Vectors), one
finds
$$
\xi=
\frac{1}{{c_2}-{c_1}}
\sqrt{\frac{m\omega}\hbar}\ \frac{\Gamma\left(({c_1}-\lambda_n)/2\right)}
{\Gamma\left(({c_2}-\lambda_n)/2\right)}
\frac{\Gamma(c_2 )}{\Gamma(c_1)},
\eqn\propfact
$$
and the spectrum $\{E_n = \lambda_n\hbar \omega \}$
is determined by the same condition as in (\spesol).
Note that, again, the spectrum depends only on the two angle parameters
$(\theta_{+},\theta_{-})$.

At some points of the angles, the spectrum becomes particularly simple.
For instance, at $(\theta_{+},\theta_{-})=(0,0)$ ({\it i.e.}, $U = I$),
we obtain
$E_n = (2n+{c_2})\hbar\omega$ and that the eigenstates are given by
$\phi_n^{(2)}(\vert x\vert)$ both on the positive and negative half lines
(and hence
each
level is doubly degenerate).  Similarly, at
$(\theta_{+},\theta_{-})=(\pi,\pi)$ ({\it i.e.}, $U = -I$), we find
$E_n = (2n+{c_1})\hbar\omega$ and that the eigenstates are
$\,\phi_n^{(1)}(\vert x\vert)\,$ which are also doubly degenerate.
As mentioned earlier, this corresponds to the Friedrichs extension and has
been conventionally considered for the quantization of the system
since Calogero [\Calogero].
On the other hand, at
$(\theta_{+},\theta_{-})=(0,\pi)$, then we have two series of eigenstates,
one
with ${N_{\rm R}^{(2)}}={N_{\rm L}^{(2)}}=0$ and the other with
${N_{\rm R}^{(1)}}={N_{\rm L}^{(1)}}=0$,
with eigenvalues
$$
E^{(1)}_n =(2n+{1 + a})\hbar\omega, \qquad
E^{(2)}_n =(2n+{1 - a})\hbar\omega, \qquad n = 0, 1, \ldots,
\eqn\evalues
$$
respectively.  In particular, in the
limit $g \to 0$ we have $a \to 1/2$, which shows that
our system recovers the spectrum of
a harmonic oscillator.  A complete reduction to the harmonic oscillator
system is realized by choosing $U = \sigma_1$, where the eigenstates
become $e^{-y^2/2}$ times the familiar
Hermite polynomials (for a detailed discussion on the smooth limit to the
harmonic
oscillator, see [\MT]).  In this respect, the extension provided by
$U = \sigma_1$ causes no obstacle at the singularity and is
called \lq the free case\rq{} in the analysis of point interactions.

\vfill\eject
\centerline
{\bf 4. Spectral space for parity invariant singular potentials}
\bigskip

The previous two examples
share the property that the spectrum of the Hamiltonian is dependent only on
the two
parameters $(\theta_+, \theta_-)$ which are determined by the eigenvalues of
the
characteristic matrix $U \in U(2)$.  This has been observed also for point
interaction [\Moebius], and can be shown to hold for any singular
potential
$V(x)$ characterized by $U(2)$, as long as it is parity invariant
$V(-x) = V(x)$.  Indeed, we have the
following
\smallskip
\noindent
{\bf Theorem.}
{\it If the Schr{\"o}dinger operator $H$ on $\Line\!\setminus\!\{0\}$
has a (measurable and locally integrable) parity invariant potential $V(x)
= V(-x)$, and is in the \lc case at $x = 0$ and in the \lp case for
$\vert x \vert \to \infty$, then its spectrum on a self-adjoint domain
${\cal D}_U$ is uniquely determined by the eigenvalues of the
characteristic matrix $U \in U(2)$.}
\smallskip

\noindent
{\it Proof.} The proof is done simply by putting the argument of the
examples in the general context.  Let $\psi_n \in L^2(\Line)$ be a
normalizable
solution of the Schr{\"o}dinger equation (\egneq) with eigenvalue $E_n$,
which is subject to the boundary condition (\BoundaryCondition) specified
by the matrix $U$.  Let also $\{\phi^{(s)}_n(x)\}_{s = 1, 2}^{}$ be a
fundamental system of real solutions for $x > 0$ with the same $E_n$.
These solutions are not necessarily subject to (\BoundaryCondition), and
are chosen to satisfy $W[\phi^{(1)}_n, \phi^{(2)}_n] = 1$.
In terms of these, the general solution of $\psi_n$ can be given
in the form (\solutions) because of the parity invariance of the
potential, $V(-x) = V(x)$.  Since (at least one but generically both of)
the basis solutions $\phi^{(s)}_n(x)$ become divergent as $x \to \infty$
as dictated by the uniqueness of the solution at the \lp infinity, one
needs to arrange the coefficients, $N_{\rm R}^{(s)}$ and $N_{\rm
L}^{(s)}$, so that the divergence of the two terms cancel each other in
the limits $x \to \pm \infty$. {}From this one deduces the equality of the
two ratios $N_{\rm R}^{(1)} : N_{\rm R}^{(2)} = N_{\rm L}^{(1)} : N_{\rm
L}^{(2)}$, that is,
$$
N_{\rm R}^{(2)} = \alpha\, N_{\rm R}^{(1)}, \qquad
N_{\rm L}^{(2)} = \alpha\, N_{\rm L}^{(1)},
\eqn\rtos
$$
with some $\alpha \in \R \cup \{\infty\}$. We also
define our reference modes (\zeromode) as
$$
\eqalign{
\varphi^{(1)}(x) &= \phi^{(1)}(\vert x\vert)\left[\Theta(x) -
\Theta(-x)\right],\cr
\varphi^{(2)}(x) &= \phi^{(2)}(\vert x\vert),
}
\eqn\genzmodes
$$
using two real eigenmodes $\phi^{(1)}$, $\phi^{(2)}$ for $x > 0$
corresponding to an arbitrary eigenvalue $E$, satisfying
$W[\phi^{(1)}, \phi^{(2)}] = 1$. Then, one obtains the following relations
for the Wronskians:
$$
W[\phi^{(s)}_n, \varphi^{(1)}]_{+0} = W[\phi^{(s)}_n, \varphi^{(1)}]_{-0},
\qquad
W[\phi^{(s)}_n, \varphi^{(2)}]_{+0} = - W[\phi^{(s)}_n,
\varphi^{(2)}]_{-0}.
\eqn\wronskrel
$$
With the help of these relations (\wronskrel) and (\rtos), one can compute
the boundary vectors (\boundvect) to find
$$
\eqalign{
\Psi
&= \left(
W[\phi^{(1)}_n, \varphi^{(1)}]_{+0} + \alpha\, W[\phi^{(2)}_n,
\varphi^{(1)}]_{+0}\right)
\pmatrix{ N_{\rm R}^{(1)} \cr
N_{\rm L}^{(1)}
}, \cr
\Psi'
&= \left(
W[\phi^{(1)}_n, \varphi^{(2)}]_{+0} + \alpha\, W[\phi^{(2)}_n,
\varphi^{(2)}]_{+0}\right)
\pmatrix{ N_{\rm R}^{(1)} \cr
N_{\rm L}^{(1)}
}.
}
\eqn\bvec
$$
One thus sees that the two complex boundary vectors are proportional to
each other, $\Psi' = \xi \, \Psi$, with a constant $\xi$ that is
specified by (\rtos) and (\bvec).  Once this is established, the rest of
the argument is already given in the examples. Namely, the boundary
condition (\BoundaryCondition) now becomes (\boundarycon)  and, hence, a
nontrivial solution is obtained if (\BoundaryConditionTwo) is fulfilled.
In terms of the parameters (\sclp)  the spectrum condition reads
(\spesol).  This proves the statement of the theorem for the bound states,
since the diagonal slots of $D$ are nothing but the eigenvalues of the
matrix $U$. In the end, we recall the fact that the continuous spectrum is
independent of $U$, since all self-adjoint extensions of a symmetric
operator admit the same continuous spectrum [\AG]. {\it Q.E.D.}
\smallskip

We note that the space of possible spectra is therefore the space of the
eigenvalues of $U$, which is $U(1) \times U(1)/\Z_2$ (where $\Z_2$ is the
factor of interchanging the two eigenvalues) forming a M{\"o}bius strip
with boundary [\Moebius].  This theorem implies that, since the separated
subfamily $\Omega_3$ contains all possible diagonal $U$, the probability
flow through the singularity plays no role as long as the spectra of
parity invariant systems are concerned.  In other words, the variety of
the spectra is exhausted by systems consisting of two separated half
lines, when all possible conditions at the boundary, {\it i.e.,}
the $U(1)$ family of boundary 
conditions each, are allowed on both sides.

\vfill\eject
\centerline
{\bf 5. Summary and Discussions}
\bigskip

In this paper we presented the generalized
connection conditions for singular potentials characterized by the matrix
$U \in U(2)$ in
the form (\BoundaryCondition) with improved boundary vectors (\boundvect).
An essential point in our generalization
is the use of Wronskians in the boundary vectors (\boundvect)
which are well-defined even in the limits to the singularity $x = 0$, in
contrast to the earlier ones (\vectors) which become ill-defined in
the limits.
Using the generalized
connection conditions, we examined two models, the one dimensional hydrogen
atom and the harmonic oscillator with inverse square potential, which are
solvable and yet so far have yielded conflicting results in the
spectrum.  Our analysis shows that the spectrum varies according to the
connection conditions adopted, and that the possible spectra form a 2-parameter
subspace (M{\"o}bius strip with boundary) in the entire
$U(2)$ family.  The confusion on the spectrum is therefore resolved once we
understand which connection conditions --- if formulated in the form
(\BoundaryCondition) --- one is using in
the analysis.   We also note that in our
connection conditions the
parameters in the matrix $U$, when combined with the decomposition (\sdecp),
bear direct physical meanings [\Moebius].  Indeed, we have already seen
this in
the spectral theorem in sect.4 in that the two parameters in the diagonal
piece $D$ represent the two independent scales of the system.

As a final remark, we wish to mention that the whole prescription for the
connection conditions remains valid even for systems with a \lq black
box\rq{}, not just a singular point. Namely, if there is a blank interval
$\I = [-\varepsilon_0, \varepsilon_0]$ with some small $\varepsilon_0 > 0$
on a line, then its quantum mechanical description can be given by means
of our prescription if one replaces $\Line\!\setminus\!\{0\}$ with
$\Line\!\setminus\! \I$, and the relevant Wronskians to be used in the
boundary vectors are $W[\phi^*, \psi ]_{\pm\varepsilon_0}$ instead of
$W[\phi^*, \psi ]_{\pm0}$. This seemingly innocent modification has a
practical consequence, since this allows us to introduce a rich $U(2)$
structure to strong singularities that are in the \lp case and are therefore
originally essentially self-adjoint with no ambiguity in connection
conditions. The appearance of the four-parameter $U(2)$ freedom in
choosing a singular potential may provide a useful theoretical framework
for describing such singularities in quantum phenomena that seem to belong
to the limit-point case by their behavior at intermediate length scales
but have a richer structure at very short length scales.
We mention that several recent approaches address the question of how to
introduce nontrivial structure to \lp singularities, see, {\it e.g.,}
[\KW].

\bigskip
\noindent
{\bf Acknowledgement:}
I.T. is indebted to H. Miyazaki for useful comments.
T.C.~thanks members of the Theory Group of KEK for
the hospitality extended to him during his stay.
This work has been supported in part by
the Grant-in-Aid for Scientific Research (C)
(Nos.~10640301 and 13640413) and that on Priority Areas (No.~13135206)
by the Japanese
Ministry of
Education, Science, Sports and Culture.

\baselineskip= 15.5pt plus 1pt minus 1pt
\parskip=5pt plus 1pt minus 1pt
\tolerance 8000
\vfill\eject\immediate\closeout\reffile
\centerline{{\bf References}}\bigskip\frenchspacing%
\input refs.tmp\vfill\eject\nonfrenchspacing

\bye